%% file: paper.tex
\documentclass[journal]{IEEEtran}
\usepackage{orcidlink}
\usepackage{cite}
\usepackage{pgfplots}
\usepackage{multirow}
\usepackage{svg}

\pgfplotsset{compat=1.18}
\def\BibTeX{{\rm B\kern-.05em{\sc i\kern-.025em b}\kern-.08em
    T\kern-.1667em\lower.7ex\hbox{E}\kern-.125emX}}
\usepackage{pifont}

\newcommand{\yes}{\ding{51}}%
\newcommand{\no}{\ding{55}}%

\begin{document}
\bstctlcite{IEEEexample:BSTcontrol}

\title{Physical Design Exploration of a Wire-Friendly Domain-Specific Processor for Angstrom-Era Nodes}

\author{Lorenzo~Ruotolo\textsuperscript{\orcidlink{0009-0004-7224-5936}},
        Lara~Orlandic\textsuperscript{\orcidlink{0000-0002-4078-7528}},
        Pengbo~Yu\textsuperscript{\orcidlink{0009-0004-8221-8608}},
        Moritz~Brunion\textsuperscript{\orcidlink{0000-0001-7842-7774}},
        Daniele~Jahier~Pagliari\textsuperscript{\orcidlink{0000-0002-2872-7071}},
        Dwaipayan~Biswas\textsuperscript{\orcidlink{0000-0001-7912-3692}},
        Giovanni~Ansaloni\textsuperscript{\orcidlink{0000-0002-8940-3775}},
        David~Atienza\textsuperscript{\orcidlink{0000-0001-9536-4947}},
        Julien~Ryckaert,
        Francky~Catthoor\textsuperscript{\orcidlink{0000-0002-3599-8515}},
        and Yukai~Chen\textsuperscript{\orcidlink{0000-0003-3378-887X}}

\thanks{L. Ruotolo and D.J. Pagliari are with the Politecnico di Torino, Turin, Italy.}
\thanks{L. Orlandic, P.B. Yu, G. Ansaloni, D. Atienza are with École Polytechnique Fédérale de Lausanne (EPFL), 1015 Lausanne, Switzerland.}
\thanks{F. Catthoor is with the National Technical University of Athens, Athens, Greece. Email: catthoor@microlab.ntua.gr} 
\thanks{L. Ruotolo, M. Brunion, D. Biswas, J. Rychaert, Y.K. Chen are with IMEC, Leuven, Belgium. Email: yukai.chen@imec.be}
}

\maketitle

\input{text/0_abstract}


\IEEEpeerreviewmaketitle

\input{text/1_intro}
\input{text/2_background}
\input{text/3_methods}
\input{text/4_results}
\input{text/5_conclusions}

\ifCLASSOPTIONcaptionsoff
  \newpage
\fi

\bibliographystyle{IEEEtran}
\bibliography{bstctl, bibliography} 

\end{document}

%% file: text/0_abstract.tex
\begin{abstract}
This paper presents the physical design exploration of a domain-specific processor (DSIP) architecture targeted at machine learning (ML), addressing the challenges of interconnect efficiency in advanced Angstrom-era technologies. The design emphasizes reduced wire length and high core density by utilizing specialized memory structures and SIMD (Single Instruction, Multiple Data) units. Five configurations are synthesized and evaluated using IMEC’s A10 nanosheet node. Key physical design metrics are compared across configurations and against VWR2A, a state-of-the-art (SoA) DSIP baseline. Results show that our architecture achieves over $2\times$ lower normalized wire length and more than $3\times$ higher density than the SoA, with low variability in the metrics across all configurations, making it a promising solution for next-generation DSIP designs. These improvements are achieved with minimal manual layout intervention, demonstrating the architecture’s intrinsic physical efficiency and potential for low-cost wire-friendly implementation.
\end{abstract}

%% file: text/1_intro.tex
\section{Introduction}

\IEEEPARstart{F}{ields} such as machine learning (ML) and digital signal processing (DSP) are rapidly evolving, consequently increasing the complexity of their applications~\cite{survey}. This trend has led to the growing adoption of custom accelerators capable of speeding up specific tasks by executing kernels on dedicated, optimized hardware, especially at the edge. While these accelerators are highly efficient compared to general-purpose CPUs, they lack flexibility and typically accelerate only a subset of workloads.

Domain-specific instruction-set processors (DSIPs) offer a more flexible alternative by supporting a broader instruction set targeted to a specific domain, enabling the acceleration of entire applications within their target field. By narrowing their scope, they represent a good compromise between the flexibility of general-purpose CPUs and the efficiency of application-specific accelerators. Several DSIP architectures have been proposed in recent years. VWR2A~\cite{vwr2a} and R-Blocks~\cite{related_rblocks} are coarse-grained reconfigurable array (CGRA) designs that offer flexibility but suffer from routing inefficiencies due to systolic array-like interconnects. Similarly, AraXL~\cite{related_araxl}, a scalable vector processor, suffers from routing congestion as its 2D block-based layout does not scale efficiently.

These limitations become increasingly critical in advanced technology nodes. As traditional scaling laws, such as Dennard scaling and Moore's Law, continue to break down, new microarchitectural solutions are required to sustain progress into the Angstrom era, where feature sizes fall below one nanometer. Although transistors continue to shrink, interconnect wires fail to scale proportionally~\cite{tokei}. While process-level mitigations have been explored through the adoption of new materials for wires and dielectrics, addressing the routing bottleneck also necessitates architectural-level solutions that are ``wire-friendly''. 
Reducing average wire length not only decreases capacitive loading, thereby lowering peak power and total energy consumption, but also reduces resistance, which helps mitigate IR drop. Additionally, thermal behavior benefits from this reduction, as power and temperature are tightly coupled~\cite{inter2}

These reductions in average wire length must be achieved without compromising scalability, as measured by computational density, in order to maximize performance per unit area. Existing DSIPs for AI or DSP workloads, such as VWR2A, although implemented in sub-nanometer nodes, only partially adopt these architectural strategies and therefore fail to exploit the potential of advanced technology fully. Addressing this limitation requires rethinking architectural design under simultaneous constraints of routing efficiency and density.

To this end, we revisit DSIP architectures with a focus on both interconnect efficiency and computational density. Motivated by these challenges, we adopt the baseline design from~\cite{jordi},
which integrates efficient scratchpad memories (SPMs) and very wide registers (VWRs)~\cite{vwr}, with flexible single-instruction multiple-data (Soft-SIMD) vector functional units (VFUs)~\cite{softsimd,catthoor2010}. The primary contribution of this work lies in the physical design exploration and comparative evaluation against similar ML-oriented DSIPs. 
We demonstrate consistent improvements across key design metrics, achieving up to 50\% reduction in routing wire length and a $3\times$ increase in core density, validating our approach on IMEC's predictive A10 (1nm) nanosheet technology node.

%% file: text/2_background.tex
\section{Background and Related Work}\label{sec:back}

The DSIP architecture employed in this work integrates several key architectural components, each embodying innovative features. This section briefly introduces them to provide the necessary background for understanding the physical design exploration that follows.


\subsubsection{Very Wide Registers}\label{sec:vwr}

Originally introduced in~\cite{vwr}, the very wide register (VWR) is a memory organization designed to achieve energy efficiency improvements over traditional register file architectures. 
Positioned close to functional units, VWRs act as foreground memories, storing data for inner-loop computations. This reduces the frequency of costly accesses to higher-level memory hierarchies, thereby minimizing data movement and enhancing locality, particularly advantageous for ML workloads that benefit from high data reuse.

Unlike multi-ported register files, where each cell connects to multiple bitlines and wordlines, each VWR cell is connected to a single bitline and a single wordline. This simplified structure allows for faster and more energy-efficient parallel access. VWRs feature a single port with two asymmetric interfaces: a wide interface for memory transfers and a narrow one for datapath operations. When paired with SIMD functional units, each word retrieved from the VWR includes multiple subwords, reducing access overhead while maintaining parallelism. 
Once the buffered data is consumed, new data is fetched from higher memory at a higher cost. Although VWRs lack the flexibility of conventional register files, being unable to perform multiple random accesses to individual subwords, this limitation is inconsequential for workloads with regular memory access patterns, such as ML inference or DSP kernels.

Multiple studies have demonstrated the energy efficiency of VWR-based architectures. In~\cite{vwr}, VWRs achieved up to $10\times$ energy savings over clustered, multi-ported register files on DSP benchmarks. The VWR2A architecture~\cite{vwr2a} demonstrated further efficiency gains, reducing system-level energy consumption by 70.8\% over the ARM Cortex-M4 and 74.4\% over the Ibex core, primarily due to its optimized internal memory structure involving VWRs, reducing expensive SoC interconnects such as AMBA AHB \cite{denkinger2023}.

\subsubsection{SoftSIMD}\label{sec:softsimd}

Traditional SIMD architectures are implemented through what can be defined as hardware SIMD (hard-SIMD), supporting only a limited set of SIMD bitwidths. This approach restricts flexibility and limits potential performance gains in applications that can benefit from fine-grained or smaller bitwidths parallelism, such as quantized ML models. 

In contrast, this work employs Soft-SIMD (Software-defined SIMD) VFUs to enhance the flexibility and efficiency of data-parallel workloads. Originally introduced in~\cite{softsimd}, Soft-SIMD enables runtime reconfiguration of SIMD widths, allowing more precise alignment with application-level parallelism requirements. The Soft-SIMD VFUs used in this design also incorporate Canonical Signed Digit (CSD) encoding, facilitating shift-add-based multipliers. By reducing the number of non-zero digits in the operands, the number of required shift-add operations is significantly reduced, leading to lower dynamic power consumption.

As demonstrated in~\cite{softsimd}, Soft-SIMD, although possibly requiring more cycles for multiplication, significantly reduces the overall Energy-Delay-Area Product (EDAP), achieving up to 56.6\% and 72.9\% improvements over Hard-SIMD in benchmark scenarios with uniform and heterogeneous data quantization, respectively.

%% file: text/3_methods.tex
\section{Design Exploration Methodology}\label{sec:method}

\begin{figure}[!htbp]
\vspace{-0.2cm}
    \centering
    \includegraphics[width=0.95\linewidth]{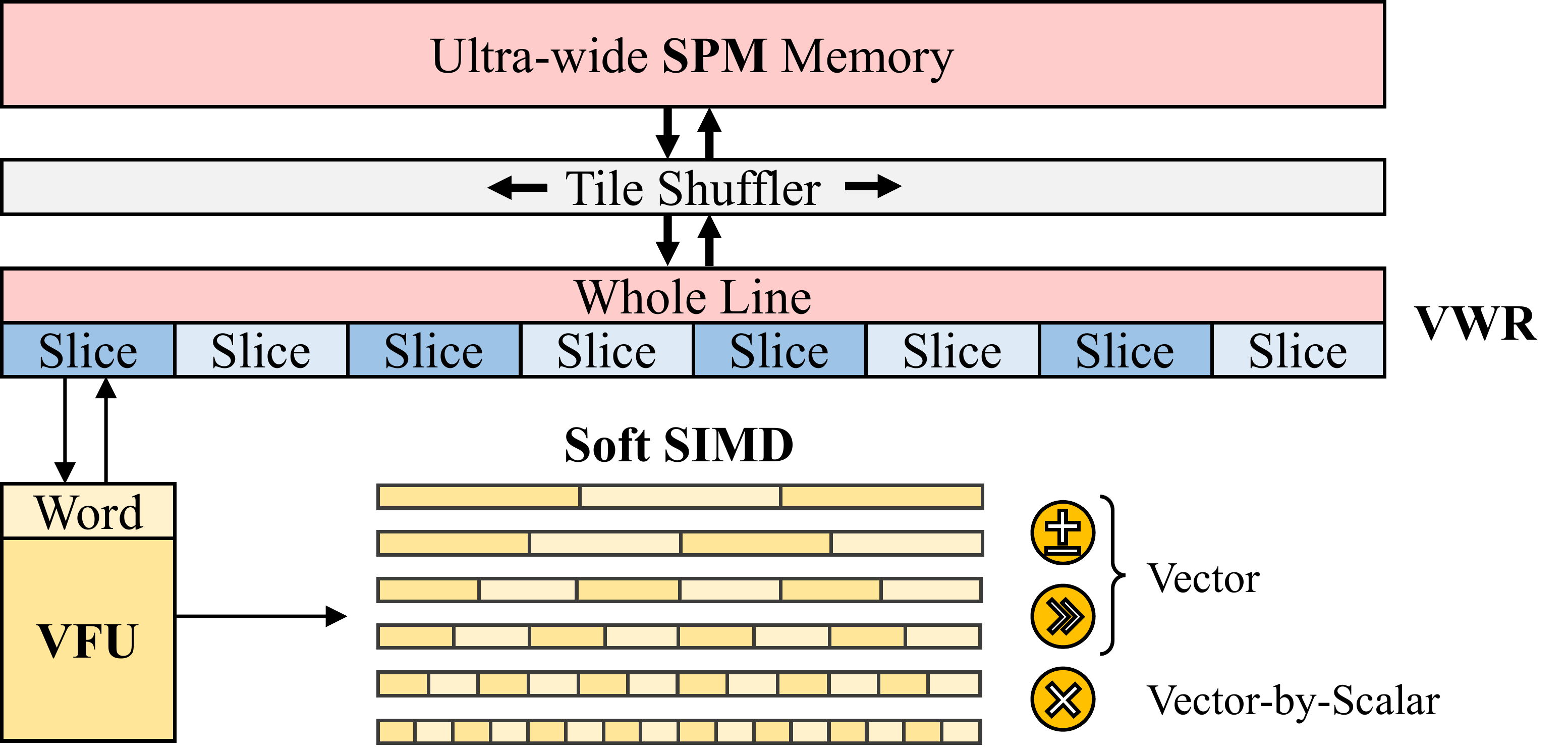}
    \vspace{-0.25cm}
    \caption{Simplified internal tile floorplan highlighting the intended layout.}
    \label{fig:tile}
\vspace{-0.2cm}
\end{figure}

The DSIP design investigated in this work adopts a tile-based architecture following the ProVeT (Processor Vector Tile) template, a specialized subset of the AERO framework~\cite{provet, catthoor2010}. In this architecture, the acceleration of data-intensive workloads is handled by specialized units called \textbf{tiles}, which can be instantiated in a 2D array to enable coarse-grained parallelism across an SoC.  A schematic representation of a single tile is shown in Figure~\ref{fig:tile}. This study focuses on the physical design exploration of an individual DSIP tile. The control plane, consistent with the implementation in~\cite{jordi}, is treated as an external component and excluded from the exploration. The ProVeT-based tile template is highly configurable, with tunable parameters such as the number and size of VWRs and VFUs, and their bitwidths, facilitating a broad and flexible design space exploration.

In terms of physical design exploration, we target a high-level floorplan where, inside each tile, the Soft-SIMD VFUs are directly interfaced with the lowest level of the system's memory hierarchy (i.e., the VWRs and SPM) through direct interconnections, similarly to other Compute-near-Memory (CnM) architectural approaches~\cite{cnm}. This enables the typical efficiency benefits of CnM designs, which stem from reduced data movement and improved locality. 

The use of simplified interconnect structures also yields concrete advantages in physical implementation. During the placement stage, shorter interconnects traversing fewer metal layers allow for tighter standard cell packing and higher achievable density. During routing, reduced wire complexity decreases routing congestion and shortens design closure times. These physical benefits would be difficult to realize in conventional designs with multi-ported register files or crossbar-based interconnects.

This section details the three primary components of the tile design: the \textbf{memory hierarchy}, the \textbf{computational units}, and the \textbf{interconnect} between them. 


\subsubsection{Memory Hierarchy}\label{sec:tile-memory}

The memory hierarchy in the proposed architecture is organized into three levels: SPM, VWRs, and local VFU registers.

The SPM serves as the primary data interface with other tiles in the array and with the system's main memory, accessed via the external Network-on-Chip (NoC). Internally, the SPM consists of SRAM banks, controlled in parallel through shared control signals. By broadcasting the same address to all banks, a common row is accessed simultaneously, forming a single N-bit wide, M-word deep L1 SPM.

At the next level, the VWRs, introduced in Sec.~\ref{sec:back}, act as L0 buffers between the SPM and the local VFU registers. Each VWR is N-bits wide, matching the width of an SPM line, and 1 bit deep, implemented with standard latch cells. VWRs are logically partitioned into words matching the VFU datapath width and grouped into slices, with each VFU connected to one slice. This allows direct access only to words in the assigned slice, reducing interface complexity and exploiting spatial locality.

If required, access to words outside a VFU’s slice is possible indirectly via slower data rearrangement using either a system-level Direct Memory Access (DMA) controller or a dedicated tile shuffler. The tile shuffler employs a one-word left shifter to enable fast reconfiguration within the VWRs. However, its area overhead scales with supported rearrangement modes, potentially reducing the space for direct VWR-SPM-VFU connections. This tradeoff must be evaluated based on workload locality.

The final memory level consists of local registers in each VFU. These one-word wide registers store intermediate operands and results. Their content is interpreted dynamically according to the active subword configuration, in line with the Soft-SIMD model: each word holds multiple subwords, enabling fine-grained parallelism.


\subsubsection{Computational Units}

The computational units in each tile implement the Soft-SIMD paradigm described in Sec.~\ref{sec:back}. Each tile contains a configurable number of VFUs with parameterizable datapath widths. While wider datapaths offer increased parallelism, they also introduce higher internal delays (primarily through the ALU’s carry chain) and greater area overhead. Consistent with the rest of the layout, VFU placement aims to minimize horizontal data-flow lengths. Less active components, such as the Data Pack Unit (DPU), are placed toward the far right of the tile to avoid interference with critical paths. A schematic of each VFU is shown in Fig.~\ref{fig:tile}, with its functionalities detailed in \cite{softsimd}. 


\subsubsection{Interconnects}

Each SPM sense amplifier output connects directly to the corresponding VWR latch interface, which shares the same physical port as the aligned VFU. This layout minimizes horizontal data-path lengths, the dominant routing direction in the tile, thereby improving locality and reducing routing complexity. Compared to traditional register files, the absence of complex multiplexing structures allows for short, bit-level, point-to-point connections between components. The most wire-efficient configuration utilizes no tile shuffler, a single VWR, and one word per slice, allowing for fully direct connections without intermediate routing or logic overhead.

%% file: text/4_results.tex
\section{Experiments}\label{sec:res}
To demonstrate the wire-efficient characteristics of the proposed design across different architectural configurations, we explore five variants (A–E) by varying the number and size of key components: SPMs, VWRs, and VFUs. 
For each configuration, a standard digital design flow was completed using the Cadence suite~\cite{cadence_genus}\cite{cadence_innovus}.
To ensure consistency, the floorplan area is set to the minimum routable area as determined by the placement tool. All designs are implemented using IMEC’s A10 predictive PDK, which represents an Angstrom-era, deeply scaled nanosheet Gate-All-Around FET (GAAFET) technology node.

To assess the benefits of our approach, we compare it against VWR2A~\cite{vwr2a}, a reconfigurable DSIP based on the CGRA paradigm. Like our design, it aims at high computational density and low-power operation, targeting applications in ML and DSP. It also incorporates energy-efficient memory components, such as SPMs and VWRs, which are also integral to our architecture. 
However, the use of traditional systolic-array-style processing elements (PE) interconnections and complex tile shuffling procedures leads to increased routing complexity, resulting in higher area and wire length requirements. 
For a meaningful and fair comparison, we synthesized a version of VWR2A using the same A10 nanosheet PDK, ensuring both designs were evaluated under identical technology constraints. Among the five generated configurations, configuration (E) most closely matches VWR2A in terms of aggregate memory sizes (24 KiB vs. 32 KiB SPM and 2304 KiB vs. 3072 KiB VWRs) and logic cell count after synthesis (304K vs. 328K), allowing for a direct architectural and physical-level comparison.

The key architectural parameters for all five configurations, along with the VWR2A baseline, are summarized in Table~\ref{tab:configs}. To facilitate the comparison, we also report the aggregate memory size in bytes for each unit, computed as the total capacity across all instances of that unit.

\renewcommand{\arraystretch}{1.15}
\vspace{-0.2cm}
\begin{table}[!h]
    \centering
    \caption{Summary of Tile Parameters Across Configurations}
    \label{tab:configs}
    \resizebox{\linewidth}{!}{
    \begin{tabular}{|c|l|r|r|r|r|r|r|}
        \hline
        \textbf{Unit} & \textbf{Parameter} & \textbf{A} & \textbf{B} & \textbf{C} & \textbf{D} & \textbf{E} & \textbf{VWR2A} \\ \hline\hline
        \multirow{3}{*}{} & Columns & 1 & 1 & 1 & 1 & 1 & 2 \\ \cline{2-8}
         & Word Width [Bits] & 96 & 192 & 96 & 192 & 192 & 32 \\ \cline{2-8}
         & Tile Shuffler & \no & \no & \no & \yes & \yes & \yes \\ \hline\hline

        \multirow{4}{*}{\textbf{SPM}} & Bank Size [Bits] & \multicolumn{6}{c|}{512$\times$64} \\ \cline{2-8}
         & Banks Number & 3 & 6 & 6 & 3 & 6 & 8 \\ \cline{2-8}
         & Bitwidth & 1536 & 3072 & 3072 & 1536 & 3072 & 4096 \\ \cline{2-8}
         & Aggregate Size [KiB] & 12 & 24 & 24 & 12 & 24 & 32 \\ \hline\hline

        \multirow{6}{*}{\textbf{VWRs}} & Number & 1 & 4 & 2 & 2 & 6 & 6 \\ \cline{2-8}
         & Bitwidth & 1536 & 3072 & 3072 & 1536 & 3072 & 4096 \\ \cline{2-8}
         & Slices Per VWR & 8 & 1 & 8 & 8 & 16 & 8 \\ \cline{2-8}
         & Words Per Slice & 2 & 16 & 4 & 1 & 1 & 32 \\ \cline{2-8}
         & Words Per VWR & 16 & 16 & 32 & 8 & 16 & 128 \\ \cline{2-8}
         & Aggregate Size [B] & 188 & 1536 & 750 & 375 & 2304 & 3072 \\ \hline\hline

        \multirow{3}{*}{\textbf{VFUs}} & Number & 8 & 1 & 8 & 8 & 16 & 8 \\ \cline{2-8}
         & Datapath Bitwidth & 96 & 192 & 96 & 192 & 192 & 32 \\ \cline{2-8}
         & Aggregate Size [B] & 96 & 24 & 96 & 192 & 384 & 32 \\ \hline
    \end{tabular}
    }
    \vspace{-0.3cm}
    \vspace{-0.2cm}
\end{table}
\renewcommand{\arraystretch}{1}


\subsection{Post-Layout Results Comparison}

\begin{figure}[!htbp]
    \centering
    \includegraphics[width=0.95\linewidth]{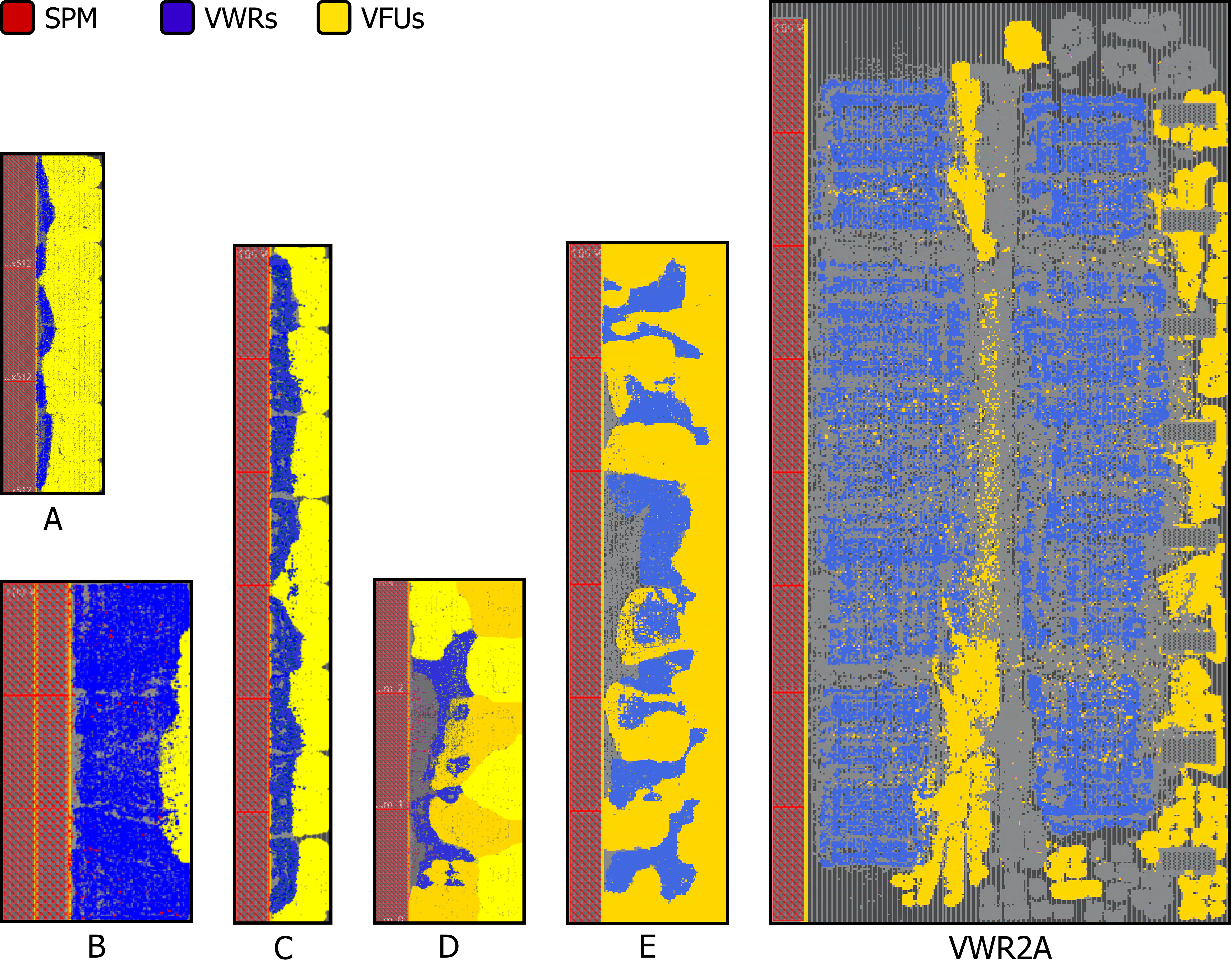}
    \caption{In-scale comparison of the physical layouts generated for configurations (A–E) and VWR2A. The SPM banks (red) are on the left side, while the standard cells on the right are grouped into VWRs (blue) and VFUs (yellow)}
    \label{fig:layout_comparison}
    \vspace{-0.1cm}
\end{figure}

The final layouts generated by the synthesis tool for the five configurations are shown in Figure~\ref{fig:layout_comparison}. Visually, it is immediately apparent that our design is significantly denser than VWR2A in all configurations, resulting in higher utilization and shorter interconnects between internal components. This observation is supported by the data presented in Table~\ref{tab:metrics_new}, which summarizes key post-layout metrics.

\begin{figure}[!htbp]
    \centering
    \includegraphics[width=\linewidth]{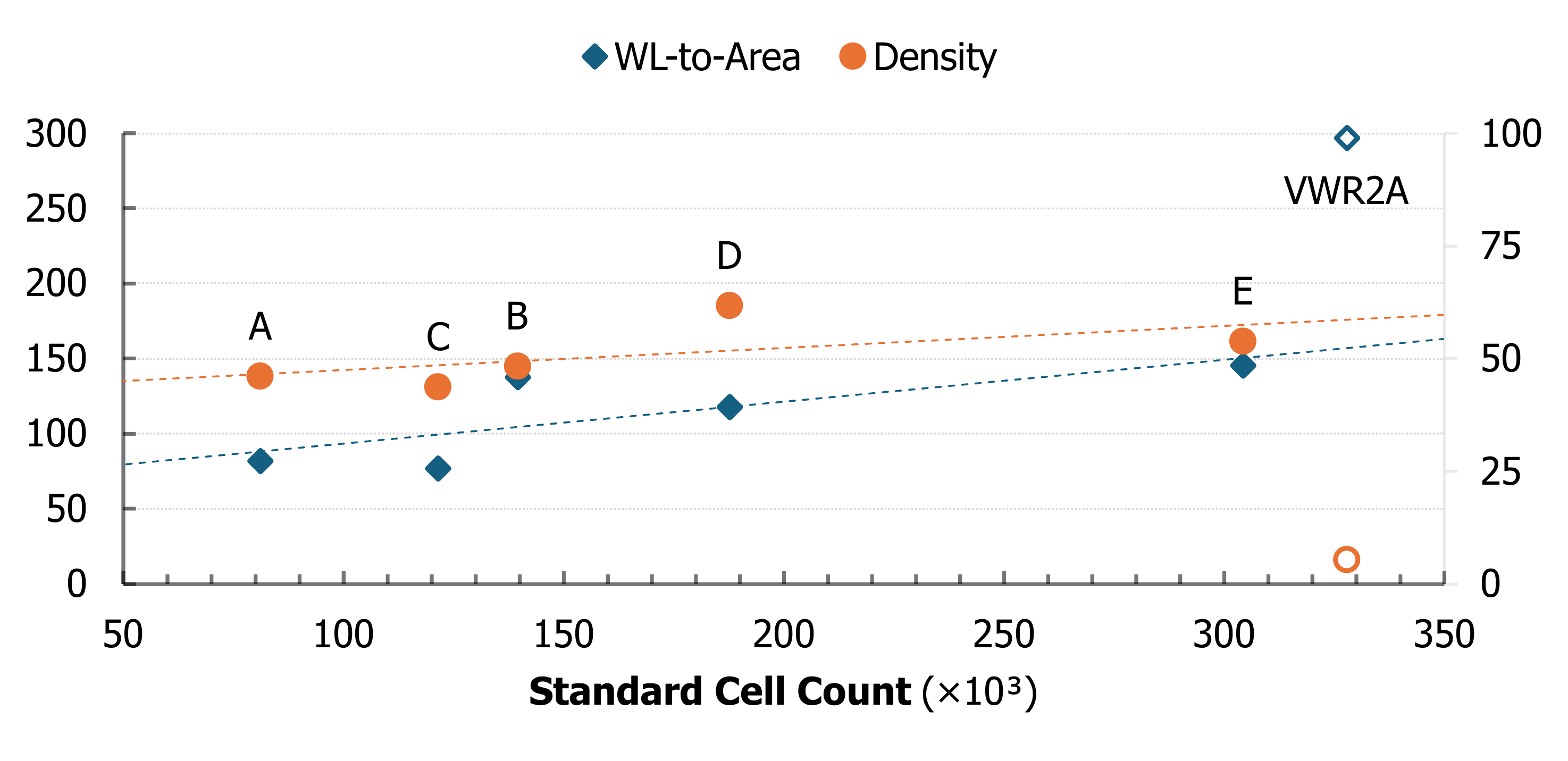}
    \caption{Plot showing trends for both WL-to-Area and core density across different logic standard cell count configurations compared to the VWR2A architecture.}
    \label{fig:trend_plot}
    \vspace{-0.2cm}
\end{figure}

The results show that, regardless of the architectural configuration, all designs achieve a core density above 40\% of the core area, with an average of 50.77\% and a standard deviation of 6.42\%. Additionally, we report the wire-length-to-area ratio as a normalized metric to evaluate wire length efficiency independently of the design size. Across all designs, this ratio averages 112.08 with a standard deviation of 28.28. These trends are visualized in Figure~\ref{fig:trend_plot}, showing the variability of the density and normalized wire length for the designs, where the horizontal axis represents the standard cell count.

When compared with the baseline of VWR2A, our selected configuration (E) has a similar number of standard cells and total logical area. Interestingly, (E) also shows no failing endpoints (FEPs) and a positive worst negative slack (WNS) of +0.004 ns in register-to-register interconnects, suggesting that this positive timing margin could allow for a higher frequency than the one used in VWR2A, potentially boosting performance. Furthermore, compared to the 16.00\% core density of VWR2A, (E) achieves 53.89\%. (E) also features a normalized interconnect wirelength that is almost $2\times$ shorter compared to VWR2A. The fact that these two metrics remain favorable and within a narrow range of variance across all configurations (A–E) shows that our design maintains the high computational density and wire-efficiency when scaling up tile configurations.

\renewcommand{\arraystretch}{1.15}
\begin{table}[h]
    \centering
    \caption{Key Metrics Across Configurations}
    \label{tab:metrics_new}
    \resizebox{\columnwidth}{!}{%
    \begin{tabular}{|l|r|r|r|r|r|r|}
        \hline
        \textbf{Metric} & \textbf{A} & \textbf{B} & \textbf{C} & \textbf{D} & \textbf{E} & \textbf{VWR2A} \\ \hline\hline
        Number of Standard Cells & 81,121 & 139,447 & 121,482 & 187,564 & 304,173 & 327,714 \\ \hline
        Total Logical Area [$\mu$m\textsuperscript{2}] & 3,372 & 6,648 & 6,092 & 5,517 & 10,632 & 15,881 \\ \hline
        reg2reg FEPs & 17 & 199 & 0 & 3335 & 0 & 114 \\ \hline
        reg2reg WNS (Setup) [ns] & -0.004 & -0.008 & +0.002 & -0.035 & +0.004 & -0.008 \\ \hline
        Wire length [$\mu$m] & 275,894 & 917,486 & 468,085 & 651,732 & 1,548,251 & 4,716,330 \\ \hline
        Wire length-To-Area Ratio & 81.82 & 138.01 & 76.84 & 118.13 & 145.62 & 296.98 \\ \hline
        Core Density & 46.09\% & 48.30\% & 43.79\% & 61.77\% & 53.89\% & 16.00\% \\ \hline
    \end{tabular}%
    }
    \vspace{-0.5cm}
\end{table}
\renewcommand{\arraystretch}{1}

%% file: text/5_conclusions.tex
\section{Conclusion}\label{sec:con}

We presented the physical design exploration of a DSIP architecture optimized for wire-efficient layouts, a key challenge in future Angstrom-era technologies. Five configurations were explored and compared to VWR2A, a similar CGRA architecture. Post-layout analysis shows that our design achieves over $2\times$ lower wire-length-to-area ratio and more than $3\times$ higher core density. Other configurations show consistent results, confirming robustness across the broad design space and demonstrating that this architectural layout is a promising solution to the energy efficiency challenges stemming from the use of advanced technology nodes for future ML/DSP DSIPs.

These improvements were achieved without guided placement of any sub-module of the design other than the SPM macro, critically reducing design effort. Future work includes evaluating such optimizations and extending the analysis to the power and thermal characteristics of the architecture.